\documentclass[aps,prd,twocolumn,showpacs,graphicx,nofootinbib]{revtex4}

\usepackage{amssymb,axodraw,graphics,graphicx}

\def\roughly#1{\mathrel{\raise.3ex\hbox
{$#1$\kern-.75em\lower1ex\hbox{$\sim$}}}}

\newcommand{\pslash}{D\kern-0.15em\raise0.17ex\llap{/}\kern0.15em\relax}
\begin{document}

\title{Hierarchically Acting Sterile Neutrinos
}
\author{Chian-Shu~Chen$^{1}$\footnote{chianshu@phys.sinica.edu.tw} and Ryo Takahashi$^{2}$\footnote{ryo.takahasi88@gmail.com}}
  \affiliation{$^{1}$Physics Division, National Center for Theoretical Sciences, Hsinchu, Taiwan 300\\$^{2}$Department of Physics, National Tsing Hua University, Hsinchu, Taiwan 300}

\date{Draft \today}
\begin{abstract}
We propose that a hierarchical spectrum of sterile neutrinos (eV, keV, $10^{13-15}$ GeV) is considered to 
as the explanations for MiniBooNE and LSND oscillation anomalies, dark matter, and baryon asymmetry 
of the universe (BAU) respectively. The scenario can also realize the smallness of active neutrino masses by seesaw mechanism. 
\end{abstract}

\maketitle
The compelling evidences from solar, atmospheric, reactor, and accelerator 
neutrino experiments have established the phenomenon of neutrino oscillations. 
The standard description is that the experimental data can be nicely explained 
by the mixings between the flavor and mass eigenstates of the three neutrinos in 
Standard Model (SM), the so-called "active" neutrinos. The unitary mixing matrix is 
parametrized in terms of three rotation angles ($\theta_{12}$, $\theta_{23}$, $\theta_{13}$) 
and one Dirac $CP$ violating phase $\delta_{CP}$. The probabilities of flavor oscillations are 
governed by the $\theta_{ij}$ and two mass-squared differences $\Delta m^2_{12} \simeq 7.59\times10^{-5}{\rm eV}^2$ 
and $|\Delta m^2_{31}| \simeq 2.45\times10^{-3}{\rm eV}^{2}$~\cite{pdg}, where $\Delta m^2_{23} >0 $ or 
$\Delta m^2_{23} < 0$ refers to normal or inverted mass hierarchy spectrum respectively. One of the 
most famous approaches to generate the active neutrino masses is the so-called "Type-I seesaw mechanism", 
in which one adds $N$ right-handed neutrinos $N_{R_{i}} (i = 1 - N)$ to the SM and the active neutrino masses can be 
obtained by block diagonalizing the mass matrix of left and right handed neutrinos, 
\begin{eqnarray}
m_{\nu_{\alpha\beta}} = - \sum_{i = 1}^{N}\frac{M_{D_{\alpha i}}M^{T}_{D_{i\beta}}}{M_{R_{i}}}. 
\end{eqnarray}
Here $\alpha, \beta = e, \mu, \tau$ represent the flavor indices of the SM fermions, $M_{D}$ is the Dirac mass 
matrix formed through the Yukawa interactions between left- and right-handed neutrinos, and $M_{R_{i}}$ 
are the Majorana masses of right-handed neutrinos. Since the right-handed neutrinos are completely neutral 
under the SM gauge symmetries, the Majorana mass $M_{R}$ is a gauge invariant quantity and $N_{R_{i}}$ 
are often termed "sterile" neutrinos. At least two sterile neutrinos are needed to accommodate the two 
mass splits observed experimentally. The mass scales of $M_{D_{\alpha i}}$ and $M_{R_{i}}$ are free parameters and 
cannot be fixed by oscillation experiments alone. 
   
There are, however, results from the LSND~\cite{Aguilar:2001ty} and the MiniBooNE~\cite{AguilarArevalo:2010wv} 
which cannot be accommodated in three active neutrinos description and may need to introduce 
one or more sterile neutrinos at the eV scale to fit the data (e.g. see~\cite{Akhmedov:2010vy,Kopp:2011qd}). It should be 
mentioned that the light sterile neutrinos are also welcome in order to have successful Big Bang 
Nucleosynthesis (BBN)~\cite{Hamann:2010bk,Hamann:2011ge}, and are consistent with the 
preference for additional relativistic degrees of freedom, $N_{eff} = 4.34^{+0.86}_{-0.88}$, observed 
from the current cosmic microwave background (CMB) anisotropy probe and the large-scale 
structure (LSS) data~\cite{pdg}.

Meanwhile, there is an increase in the amount and precision of cosmological data indicates that about $80\%$ of 
the matter content in the universe is non-baryonic dark matter (DM). The study of nature of DM is 
one of the main topics in cosmology, astrophysics, and particle physics. The Cold Dark Matter (CDM) is widely studied 
partly because the WIMP (weakly interacting massive particle) may reveal its signal at LHC and is predicted in many 
popular models (supersymmetric models, etc). However, it has been noticed that a sterile neutrino with mass at 
the keV scale and with small mixing to the active neutrinos can make up the DM in the form of 
Warm Dark Matter (WDM)~\cite{hep-ph/9303287,astro-ph/9810076}. 
Additionally, it was pointed out that the keV sterile neutrino might play an important role in explaining 
the pulsar kicks~\cite{hep-ph/9701311}.

Finally, the level of one out of ten billions excess in the amount of matter over antimatter is a long standing puzzle 
for high energy physicists. The observation hints that baryon number ($B$) and/or lepton number ($L$) are violated in 
certain physical processes. Grand unified theories (GUTs) naturally provide a framework for breaking $B$ and $L$, 
in which the fundamental fermions - quarks and leptons - are arranged in the same multiplets, and 
the out-of-equilibrium decays of heavy gauge bosons or colored Higgs bosons $H_{C}$ will generate the sufficient 
baryon asymmetry around the scale of grand unification~\cite{TU/78/179,ITP-616-STANFORD, 
Print-78-0858-Rev. (PRINCETON), HUTP-79-A050,Barr:1979ye}. It was then recognized that the Standard Model 
(SM) violates $B + L$ symmetry through the $SU(2)_{L}$ global anomaly~\cite{PRINT-76-0254 (HARVARD)}. 
The process is not suppressed during the period $100$ GeV $\lesssim T \lesssim10^{12}$ GeV, and the solution 
is called "sphalerons"~\cite{IC/85/8}. The 
sphaleron effect violates $B + L$ but conserves $B - L$, and therefore, it would erase any primordial $B + L$ 
asymmetry. We notice that any grand unification theories with higher symmetries respects $B - L$ symmetry. 
For example, $B - L$ is a global symmetry for $SU(5)$ GUTs and a local symmetry for $SO(10)$ GUTs respectively. 
The GUT-baryogenesis, therefore, is not able to explain baryon asymmetry in our universe (BAU). 
To solve the problem one has to generate $B - L$ asymmetry by violating pure baryon number~\cite{DOE-ER-01545-444} 
or by violating pure lepton number~\cite{Fukugita:1986hr} (leptogenesis, e.g.), and the sphaleron 
process will convert partially $B - L$ asymmetry into baryon asymmetry. We adopt the construction that one heavy sterile 
neutrino causes lepton asymmetry during the epoch the sphalerons are ineffective, and the late decay of colored Higgs 
will generate the observed BAU.

We consider a scenario of three sterile neutrinos $N_{R_{i(i=1-3)}}$ with hierarchical mass spectrum 
($M_{R_{1}}\sim $eV, $M_{R_{2}}\sim$keV, $M_{R_{3}}\sim10^{13-15}$ GeV), in which the lightest one may help to 
explain the neutrino oscillation anomalies, keV-scale 
sterile neutrino is the candidate of dark matter, and the heaviest state $N_{R_{3}}$ would resurrect the 
GUT-baryogenesis. Three sterile neutrinos can be introduced to cancel the additional gauge 
anomaly for any theory beyond SM with extra gauge $U(1)_{B-L}$ symmetry. We show this hierarchical spectrum 
of sterile neurinos simultaneously satisfy the observations, and how our scenario fits in the framework of GUT theories. 

It has been proposed that models of SM with (three) additional sterile neutrinos are phenomenologically 
viable~\cite{hep-ph/0503065,arXiv:1006.1731,arXiv:1110.6382}. The so-called $\nu$MSM ($\nu$ Minimal 
Standard Model)~\cite{hep-ph/0503065}, in which a mass of a keV sterile neutrino is responsible for DM, 
and two heavier states with degenerate masses lain in the range 1 GeV $\sim$ 100 GeV are required to be in 
thermal equilibrium around electroweak scale in order to generate BAU through the resonant neutrino oscillations. 
The split seesaw model with three sterile neutrinos living in the extra dimension (ED) is shown to be able to solve DM and BAU 
as well~\cite{arXiv:1006.1731}. By utilizing an exponential factor in the size of ED one can split the Majorana 
masses of $N_{R_{i}}$ with relative mild parameters associated to their locations in ED. Recently a flavor symmetry 
model~\cite{arXiv:1110.6382} proposed by Barry, Rodejohann, and Zhang (BRZ), it consists of two $N_{R_{1,2}}$ 
masses at eV-scale and one $N_{R_{3}}$ at keV-scale. The two eV-scale sterile neutrinos are used to explain LSND and 
MiniBooNE anomalies while the keV sterile neutrino is the WDM particle. The scenarios are summarized 
in Table \ref{scenario}. 
\begin{table}[t]
\caption{\label{scenario} The content of three sterile neutrinos models}
\begin{ruledtabular}
\begin{tabular}{|c|c|c|c|c|}\hline
Models & eV & keV & GeV & $\gg$EW \\
\hline
$\nu$MSM & & $N_{R_{1}}$ &$N_{R_{2}}, N_{R_{3}}$& \\ \hline 
Split Seesaw & & $N_{R_{1}}$ & & $N_{R_{2}}, N_{R_{3}}$ \\ \hline
BRZ &$N_{R_{1}}, N_{R_{2}}$&$N_{R_{3}}$& & \\ 
\hline 
HASN & $N_{R_{1}}$ & $N_{R_{2}}$ & & $N_{R_{3}}$\\ \hline  
\end{tabular}
\end{ruledtabular}
\end{table}
These setup can answer two of the three puzzles we mentioned above while 
our hierarchically acting sterile neutrinos (HASN) scenario would explain the three puzzels simultaneously. The splittings of the 
sterile neutrino masses can be achieved by implementing split seesaw mechanism~\cite{arXiv:1006.1731} or 
Froggatt-Nielsen (FN) mechanism~\cite{CERN-TH-2519} to the model. 

The Lagrangian which is relevant to neutrino masses has the form 
\begin{eqnarray}
{\cal L} &=& {\cal L}_{\rm SM} + i\bar{N_{R_{i}}}\not{\partial}N_{R_{i}} - y_{\alpha i}H^{\dag}\bar{l}_{\alpha}N_{R_{i}} \nonumber \\
&& - \frac{M_{R_{i}}}{2}\bar{N}^c_{R_{i}}N_{R_{i}} + \rm h.c. 
\end{eqnarray}
Here ${\cal L}_{\rm SM}$ is the SM Lagrangian, $l_{\alpha}$ are $SU(2)_{L}$ leptonic doublets with flavor index $\alpha$, 
$H$ is SM Higgs, $y_{\alpha i}$ are the Yukawa couplings, and $c$ is charged conjugation. The Majorana mass matrix 
of sterile neutrinos is chosen to be diagonal without loss of generality. The $6\times6$ neutrino mass matrix is given 
in the form 
\begin{eqnarray}
\left(\begin{array}{cc}0 & M_{D} \\M^{\dag}_{D} & M_{R}\end{array}\right)
\end{eqnarray}
in the basis $(\nu_{e}, \nu_{\mu}, \nu_{\tau}, N_{R_{1}}, N_{R_{2}}, N_{R_{3}})$, and $M_{R}=\rm{diag}({\cal O}(\rm eV), {\cal O}(\rm keV), {\cal O}(10^{13-15}) \rm GeV)$. Here we give a brief comment on a realization of such a hierarchical 
mass spectrum of right-handed neutrinos. One of simple examples to realize it 
is to utilize the split seesaw mechanism, in which 
the spinor fields are introduced in a flat five 
dimensional (5D) spacetime whose compactification length of extra dimension is $\ell$ and all SM 
particle are assumed to live in a 4D-brane. After solving the 5D Dirac equation and 
identifying the zero-modes of the 5D spinors with the right-handed neutrinos, the 
effective (4D) right-handed Majorana masses are described by exponential 
functions as $M_{Ri}=2\kappa_im_iv_{B-L}/(M(e^{2m_i\ell}-1))$ where 
$\kappa_i$, $m_i$, $v_{\rm B-L}$, and $M$ are a coupling constant of order one,
 bulk masses for 5D spinors, $U(1)_{{B-L}}$ breaking scale, and 5D 
fundamental scale respectively. In this mechanism, one can easily obtain a 
hierarchical right-handed neutrino mass spectrum such as 
$(M_{R_1},M_{R_2},M_{R_3})=(1\mbox{ eV},1\mbox{ keV},10^{13}\mbox{ GeV})$ 
within a set of moderate parameters when one takes $\kappa_i=1$, 
$v_{\rm B-L}=10^{15}$ GeV, and 
$(M\ell,m_1\ell,m_2\ell,m_3\ell)=(30,27.9,24.4,1.03)$ as reference values. The 
FN mechanism can also give a hierarchical mass spectrum with appropriate 
$U(1)_{{\rm FN}}$ charges.

After electroweak symmetry breaking where Higgs develops its vacuum expectation 
value (VEV) $v = 174$ GeV, one gets Dirac neutrino mass terms. The left-handed 
neutrinos receive their Majorana masses through seesaw mechanism, we obtain 
\begin{eqnarray}
m_{\nu_{3}} &\sim& 
\left\{
\begin{array}{ll}
m_{atm} \simeq \frac{|y_{\alpha3}^\ast y_{\beta3}|v^2}{M_{R_{3}}} & \mbox{for NH}
\\
\epsilon\simeq\frac{|y_{\alpha2}^\ast y_{\beta2}|v^2}{M_{R_{2}}} & \mbox{for IH}
\end{array}
\right., \label{mnu3}\\
m_{\nu_{2}} &\sim& m_{sol} \simeq \frac{|y_{\alpha1}^\ast y_{\beta1}|v^2}{M_{R_{1}}}~~~\mbox{for both NH and IH}, \\
m_{\nu_{1}} &\sim& 
\left\{
\begin{array}{ll}
\epsilon\simeq\frac{|y_{\alpha2}^\ast y_{\beta2}|v^2}{M_{R_{2}}} & \mbox{for NH} \\
m_{atm} \simeq \frac{|y_{\alpha3}^\ast y_{\beta3}|v^2}{M_{R_{3}}} & \mbox{for IH}
\end{array}
\right.  \label{mnu1}
\end{eqnarray} 
at the leading order, where NH and IH mean the normal hierarchy and inverted 
hierarchy respectively. The indices $\alpha$ and $\beta$ in $m_{\nu_3}$ for 
NH should correspond to only $\mu$ and $\tau$ in order to be consistent with 
the current data of neutrino oscillation experiments, that is, there are a 
maximal atmospheric, a large solar, and a small reactor mixing angles. By 
choosing a set of appropriate values of the Yukawa couplings, the 
experimentally observed mixing angles can be always fitted in our scenario. The
 degenerated mass spectrum of active neutrinos can be also realized. 
 
The dark matter candidate of the scenario is decaying DM. The keV sterile 
neutrino $N_{R_{2}}$ should live longer than the age of the universe and can be estimated as 
$\tau_{N_{R_{2}}} \simeq 5\times10^{26}(M_{R_{2}}/\rm{keV})^{-5}(10^{-8}/\Theta^{2})\rm{s}$, 
here $\Theta$ is the mixing between keV sterile neutrino and active neutrinos. The generic way to produce 
DM is through the active-sterile neutrino oscillations~\cite{hep-ph/9303287}, however, the abundance 
is constrained by the X-ray observations~\cite{astro-ph/0106002} (also see~\cite{arXiv:0901.0011} and 
references therein), structure formation simulations, and the Lyman-$\alpha$ bounds~\cite{astro-ph/0106108}. 
One way to relax the restrictions was proposed by Shi and Fuller~\cite{astro-ph/9810076} that an enhancement 
of the production of keV sterile neutrino can be realized via lepton-number-driven MSW 
(Mikheyev-Smirnov-Wolfenstein) effect. The other possibility is the $N_{R_{2}}$ pair production via 
$U(1)_{\rm{B-L}}$ gauge boson exchange~\cite{arXiv:1006.1731}. It has been shown that as long as the reheating temperature 
is about $10^{13}$ GeV one can account for the relic abundance of DM.    
The corresponding Yukawa couplings of sterile neutrino DM 
for the required mass $\mathcal{O}(1)$ 
keV$\lesssim M_{R_2}\lesssim\mathcal{O}(10)$ keV are typically restricted to 
$\mathcal{O}(10^{-15})\lesssim|y_{\alpha2}|\lesssim\mathcal{O}(10^{-13})$ to 
satisfy astrophysical constraints (see e.g. \cite{arXiv:0901.0011} and 
references therein). This means that terms from the sterile neutrino DM through
 the seesaw mechanism does not contribute to the atmospheric and solar neutrino
 mass scales shown in (\ref{mnu3})-(\ref{mnu1}). The Yukawa couplings for the 
1st and 3rd generations of right-handed neutrinos are approximated as
  $|y_{\alpha3}^\ast y_{\beta3}|^{1/2}\sim\mathcal{O}(0.1)$ and 
$|y_{\alpha1}^\ast y_{\beta1}|^{1/2}\sim\mathcal{O}(10^{-13}$-$10^{-12})$ to satisfy  
the atmospheric and solar scales with 
$(M_{R_1},M_{R_3})=(1\mbox{ eV},10^{13}\mbox{ GeV})$ respectively. It is seen 
that the construction of active neutrino mass spectrum in HASN scenario is 
consistent with the constraints on keV sterile neutrino DM. It can be also found that 
the atmospheric scale can be derived from the ratio of the right-handed neutrino mass, 
$M_{R_3}\sim\mathcal{O}(10^{13})$ GeV, and the corresponding Dirac masses, 
$|y_{\alpha3}^\ast y_{\beta3}|^{1/2}v\sim\mathcal{O}(10)$ GeV, when $N_{R_{3}}$ gets integrated out. 
While the solar scale comes from the seesaw relation between $M_{R_1}\sim\mathcal{O}(1)$ eV 
and $|y_{\alpha1}^\ast y_{\beta1}|^{1/2}v\sim\mathcal{O}(0.1)$ eV. Finally, the ratio of mass scales 
between the 2nd generation of sterile neutrino (DM), $M_{R_2}\sim\mathcal{O}(1)$ keV, and the 
corresponding Dirac masses, $|y_{\alpha2}^\ast y_{\beta2}|^{1/2}v\sim\mathcal{O}(10^{-3}$-$10^{-1})$ eV, 
is too steep to contribute to the active neutrino mass (atmospheric and solar) 
scales. Therefore, the sterile neutrinos are hierarchically acting also for 
giving the active neutrino mass scales. The Yukawa structure realizing 
the scenario can be obtained in both split seesaw and FN mechanisms with 
appropriate model parameters. 


Now we come to the phenomena of neutrino oscillation anomalies. The LSND 
$\bar{\nu}_{\mu} \rightarrow \bar{\nu}_{e}$ transitions anomaly reported a 3.8$\sigma$ 
excess of $\bar{\nu}_{e}$ candidate events, in which the neutrino fluxes were produced by 
dumping 800 MeV protons into a "beam stop" which mostly generate $\pi^+$, and neutrinos 
(anti-neutrinos) are the decay products of pions. The probability that $\nu_{a}$ oscillates into 
$\nu_{b}$ is given by 
$P(ab) = \sin^2(2\theta)\sin^2(1.27\Delta m^2\frac{L}{E})$,
where $\theta$ is the mixing angle, $L$ is the neutrino travel distance in the unit of meter, and 
$E$ is the neutrino energy in MeV.  The typical anti-neutrinos energies are a few MeV for reactor 
experiments, the excess is interpreted as the hints for $\bar{\nu}_{\mu} \rightarrow \bar{\nu}_{e}$ 
oscillation with $\Delta m^2 \sim 1~\rm eV^2$. This indicates at least one sterile neutrino with 
mass at the eV-scale. Then the MiniBooNE experiment set out to check the excess events in 
the $\nu_{\mu} \rightarrow \nu_{e}$ transitions and found the parameters were not compatible 
with LSND~\cite{arXiv:0704.1500}. However, more recently the MiniBooNE accumulated more 
anti-neutrino oscillation data and reported the excess electron anti-neutrino appearance is reconciled 
with LSND results~\cite{AguilarArevalo:2010wv}. To accommodate the neutrino and anti-neutrino 
data the additional CP violation has to be invoked. One simple way is to add two sterile neutrinos 
at eV scale (the so-called (3+2) scheme) to neutrino sector, the CP violation at short-baselines would 
let to reconcile both LSND and MiniBooNE results~\cite{hep-ph/0011054,hep-ph/0305255,arXiv:0705.0107,
Akhmedov:2010vy,Kopp:2011qd}. In our scenario a (3+1) scheme together with nonstandard interactions (NSI) 
of neutrinos will allow to fit the data~\cite{Akhmedov:2010vy}. The new interactions may modify the charged and 
neutral currents, and provide the new sources of CP violation. They may affect the neutrino oscillations via the 
production, propagation, and the detection processes. The four-fermion operators can be expressed at low energies as 
\begin{eqnarray}
{\cal L}_{\rm{NSI}} = 2\sqrt{2}G_{F}\sum_{f}\epsilon_{\alpha\beta}^{f_{L,R}}(\bar{\nu}_{L\alpha}\gamma^{\mu}\nu_{L\beta})
(\bar{f}_{L,R}\gamma_{\mu}f_{L,R}) + {\rm h.c.},
\end{eqnarray} 
where $G_{F}$ is Fermi constant, $f$ represents fermions (charged leptons and quarks), and $\alpha,\beta$ 
are flavor indices, and $L,R$ are chiralities. The new interactions can be induced from several possibilities 
of physics beyond SM. For example, in GUT theories the $\psi(\bold{10})\oplus\psi(\bar{\bold{5}})$ fermion 
representations of $SU(5)$ are coupled to a $\bold{5} (H_{C})$ and a $\bar{\bold{5}} (\bar{H}_{C})$ representation of Higgs, and 
the Yukawa interactions read ${\cal L}_{Y} = \psi(\bold{10})^{T}\lambda^{u}\psi(\bold{10})H_{C} + \psi(\bold{10})^{T}\lambda^{d}\psi(\bar{\bold{5}})\bar{H}_{C}$. It has been shown that one can fit to global short-baseline data for $\epsilon_{\alpha\beta}\sim{\cal O}(10^{-2})$~\cite{Akhmedov:2010vy}.    

In the context of our consideration, we adopt that a sterile neutrino $N_{R_3}$ is heavier than 
the colored Higgs bosons, and a lepton number violating interaction $l\phi l\phi/(2M_{R_3})$, 
here $\phi$ is the Higgs doublet, keeps in thermal equilibrium when $T\gtrsim10^{12}$ GeV. 
Therefore, all lepton asymmetry generated by colored Higgs decay is erased 
by the process $l+\phi\rightarrow\bar{l}+\phi^\dagger$ while the generated baryon asymmetry 
remains intact, and so $B-L\neq0$ can be satisfied.
When temperature drops below $10^{12}$ GeV the sphaleron transitions become effective, as the results, the 
produced baryon asymmetry is partially converted into the lepton asymmetry but a residual baryon asymmetry 
remains, and thus the observed BAU can be generated. In the case of $SU(5)$ GUT we have the Yukawa couplings 
$\psi(\bold{10})^{T}\lambda^{u(k)}\psi(\bold{10})H_{C}^{(k)}$ and 
$\psi(\bold{10})^{T}\lambda^{d(k)}\psi(\bar{\bold{5}})\bar{H}_{C}^{(k)}$ $(k=1,2)$. The size of BAU is calculated 
as~\cite{Barr:1979ye,Fukugita:1986hr,Fry:1980bd,Botella:1990vf}
\begin{eqnarray}
 Y_{\Delta B}&\equiv&\frac{n_B-n_{\bar{B}}}{s}=0.35\cdot0.5\cdot10^{-2}\cdot\frac{\epsilon_B}{1+(3K)^{1.2}}, 
\end{eqnarray}
where $K\equiv\frac{1}{2}\frac{\Gamma}{H}|_{T=m_{H_{C}}}\simeq\frac{1.1\times10^{18}\mbox{ GeV}}{{\lambda^{u}}^2m_{H_{C}}}\left(\frac{1}{g_\ast}\right)^{1/2}$ is the washout factor and $\epsilon_B\simeq\frac{\eta_1}{8\pi}\cdot10^{-2}[F(x)-F(1/x)+G(x)-G(1/x)]$ 
is the CP-asymmetry with $\Gamma$, $H$, $m_{H_{C}}$, $g_\ast$, $x$ are the decay rate, 
expansion rate, colored Higgs mass, degrees of freedom $g_\ast|_{T\simeq m_{H_{C}}}\simeq53$, 
mass ratio $m_{H_{C}^{(2)}}^2/m_{H_{C}^{(1)}}^2$, respectively. The functions $F$ and $G$ are defined as 
$ F(x)\simeq1-x\ln\left(\frac{1+x}{x}\right)$ and $G(x)=\frac{1}{x-1}$. The factor $0.35$ comes from the 
sphaleron process, and  
we take $\eta_1=\sin(\mbox{arg}[\mbox{tr}(\lambda^{d(1)\dagger}\lambda^{d(2)}\lambda^{u(1)\dagger}\lambda^{u(2)})])$ 
and $\eta_1\simeq\eta_2$. It is seen that we can realize $Y_{\Delta B}=8.75\times10^{-11}$ when we set $(m_{H_{C}^{(1)}},m_{H_{C}^{(2)}})=(9\times10^{12},8\times10^{12})$ GeV and $\eta_1\simeq-0.444$. 
These values are consistent with this baryogenesis scenario and above discussion of the right-handed 
neutrinos mass spectrum, that is, $10^{12}\mbox{ GeV}\leq m_{H_{C}^{(i)}}<M_{R_3}\lesssim10^{15}$ GeV. 
A heavier mass of the corresponding sterile neutrino as $10^{14-15}$ GeV is also possible for this baryogenesis.

The hierarchical spectrum of sterile neutrinos (eV, keV, $10^{13-15}$ GeV) is a simple and economical scenario, 
especially it can be embedded in many frameworks beyond SM. In light of the puzzles from neutrino oscillation 
anomalies, dark matter, and baryon asymmetry of the universe (BAU) we have shown that this scenario is 
phenomenologically viable. Searching for deviations from standard three active neutrino oscillations and the X-ray 
astronomy will offer opportunity to test the scenario.




This work has been supported in part by funds
from the National Science Council of Taiwan under Grant Nos.
NSC97-2112-M-006-004-MY3, NSC-98-2112-M-007-008-MY3 and 
the National Center for Theoretical Sciences and National Tsing Hua 
University, Taiwan.

\end{document}